\documentclass[letterpaper, 10 pt, conference]{IEEEtran}  

\IEEEoverridecommandlockouts                              

\usepackage{amsmath,amssymb,amsfonts}
\usepackage{algorithmic}
\usepackage{graphicx}
\usepackage{textcomp}
\usepackage[table]{xcolor}
\usepackage{caption}
\usepackage{multirow}
\usepackage{footnote}
\usepackage{hyperref}
\usepackage{array}
\usepackage{amssymb}
\usepackage{todonotes}
\usepackage[pscoord]{eso-pic}
\usepackage{multirow}
\usepackage{mathtools}
\usepackage{cite}

\usepackage{comment}

\usepackage{geometry}

\title{\LARGE \bf
Enhancing Train Transportation in Sri Lanka: A Smart IOT based Multi-Subsystem Approach using MQTT
}

\makeatletter

\author{Dhanushka Balasingham$^{1}$, Sadeesha Samarathunga$^{2}$, Anuththara Bandara$^{3}$, Gayantha Godakanda Arachchige$^{4}$, \\
Narmada Gamage$^{5}$ and Jaliya L. Wijayaraja$^{6}$
\\
\IEEEauthorblockA{$^{1-6}$ Dept. of Computer Systems Engineering, Sri Lanka Institute of Information Technology, Sri Lanka}

\thanks{ 
        {\tt\small $^{1}$dhanushkasandeepa@gmail.com, \tt\small $^{5}$narmada.g@sliit.lk, 
\tt\small $^{6}$jaliya.wijayaraja@gmail.com} (Correspondence author: Dhanushka Balasingham)}%
}

\begin{document}

\maketitle

\begin{abstract}
This research proposes a system as a solution for the challenges faced by Sri Lanka's historic railway system, such as scheduling delays, overcrowding, manual ticketing, and management inefficiencies. It proposes a multi-subsystem approach, incorporating GPS tracking, RFID-based e-ticketing, seat reservation, and vision-based people counting. The GPS-based real-time train tracking system performs accurately within 24 meters, with the MQTT protocol showing twice the speed of the HTTP-based system. All subsystems use the MQTT protocol to enhance efficiency, reliability, and passenger experience. The study's data and methodology demonstrate the effectiveness of these innovations in improving scheduling, passenger flow, and overall system performance, offering promising solutions for modernizing Sri Lanka's railway infrastructure.

\end{abstract}


\section{Introduction}

Sri Lanka's railway system, inaugurated in 1858 during the colonial era, has grown to connect major urban centers and diverse regions. Today, it is essential for commuter and cargo transport. However, the system faces significant efficiency challenges, including unreliable scheduling, frequent delays, and outdated infrastructure leading to breakdowns and service disruptions. Overcrowding during peak hours, limited seating, and inadequate facilities worsen the passenger experience. Manual ticketing processes are inefficient, causing long queues and boarding delays. The absence of real-time tracking and monitoring systems further hampers efficient management and rapid incident response, while the lack of precise data on passenger flow and train operations complicates resource allocation and decision-making. Modernizing the system with innovative technologies, as proposed in this research, can address these issues. However, the variety of systems from different vendors and diverse tracking platforms create interoperability challenges, complicating integration and management.
The primary aim of our research was to implement an Autonomous Train Management System (ATMS) to enhance the existing traditional transportation infrastructure through the utilization of cutting-edge technologies such as the Internet of Things (IOT) and Artificial Intelligence (AI). The key contributions of our undertaking are outlined as below:

\begin{itemize}
  \item Global Positioning System (GPS) based train tracking system with alarm notification system
  \item Radio Frequency Identification (RFID) based Electronic ticketing (E-ticketing) system
  \item An vision-based real time people counting system
    \item Proposed a standardized Message Queuing Telemetry Transport (MQTT) topic for public transportation system
\end{itemize}

\section{Related Work}
\label{sec:bgr}

The MQTT protocol is often preferred over the Hypertext Transfer Protocol (HTTP) for IoT applications. According to Tetsuya Yokotani et al. \cite{7814989}, MQTT offers advantages in terms of power consumption and speed. A comprehensive study by the authors of \cite{8394143} compared various application layer protocols used in IoT and highlighted MQTT's benefits over HTTP, Constrained Application Protocol (CoAP), Advanced Message Queuing Protocol (AMQP), and Extensible Messaging and Presence Protocol (XMPP). They emphasized MQTT's effectiveness in scenarios where an actuator network must respond to a shared sensor input.

Various vehicle tracking technologies exist \cite{7266601}, with cellular and Wi-Fi-based systems cost-effective in urban areas but impractical in rural regions due to infrastructure limitations. Previous studies \cite{8065840,9683883} proposed GPS-based vehicle tracking systems using the HTTP protocol to send location information, but this approach can be inefficient in terms of power consumption and may suffer from latency issues, resulting in non-real-time location tracking. The MQTT protocol offers a significant advantage in scalability over HTTP, making it ideal for our system, which needs to track multiple trains and accommodate numerous users simultaneously. This scalability ensures greater stability and reliability in managing various operations concurrently. By utilizing multiple Quality of Service (QoS) levels, our system minimizes bandwidth usage compared to the HTTP protocol \cite{8273112}. This optimization enhances system efficiency and performance, ensuring optimal resource utilization while delivering robust functionality.

Existing IoT-based approaches for managing transportation in Sri Lanka, as seen in \cite{9708053,9946624}, share the common objective. However, these systems operate independently without a unified backend, posing challenges for startups due to the lack of standardization. This results in lower reliability and necessitates repeated implementation efforts. In contrast, the system proposed in \cite{9501928} utilizes an MQTT-based location tracking system for theft detection, enhancing personalized usage with Google Maps integration. Our approach introduces an open-source standardized MQTT topic for vehicle tracking, enhancing versatility and interoperability across various applications and systems, benefiting vendors, users, and researchers alike.

Many ticketing systems in Sri Lanka still use traditional paper tickets, contributing to resource waste and time-consuming queues \cite{10277708}. Kazi et al. \cite{8537302} introduced an e-ticketing system for buses, delivered via bill or SMS, requiring passengers to have a mobile phone and undergo tedious ticket checks. In contrast, our RFID-based e-ticketing system allows passengers to carry a ticket or account number, accessible to all ages without a phone. Additionally, our ticket checking process is automated using RFID scanners at entrances. In \cite{9732728}, the authors proposed an RFID-based e-ticketing system where ticket pricing relies on GPS information gathered via the SIM-800L module, using the cellular network. However, this method may lead to inaccuracies in rural areas. In contrast, our system tracks trains using the NEO-6m module, which relies on GPS technology, ensuring more accurate tracking regardless of location. 

As trains traverse vast distances, they accommodate a high volume of passengers at various times.An automated passenger counting system benefits both the decision-making process of management and the passengers themselves. The authors of this research paper \cite{9391951} used Visual Geometry Group-16 (VGG-16) model for object detection to count the people. In our proposed system, You Only Look Once (YOLO) version 8 model was used to detect people. The comparison researches \cite{Chen_2024}  indicates that YOLOv8 demonstrates superior performance in terms of detection time, precision, and ease of implementation, especially considering the frames needed for real-time usage than the VGG-16 model.

\section{Methodology}
\label{sec:mtd}

The proposed system comprises software components such as the MQTT protocol, Arduino, Python, YOLO algorithms, and computer vision. The system consists of four key subsystems that work collaboratively to achieve autonomous management for train transportation.

\begin{figure}[ht]
    \centering
    \includegraphics[width=1\linewidth]{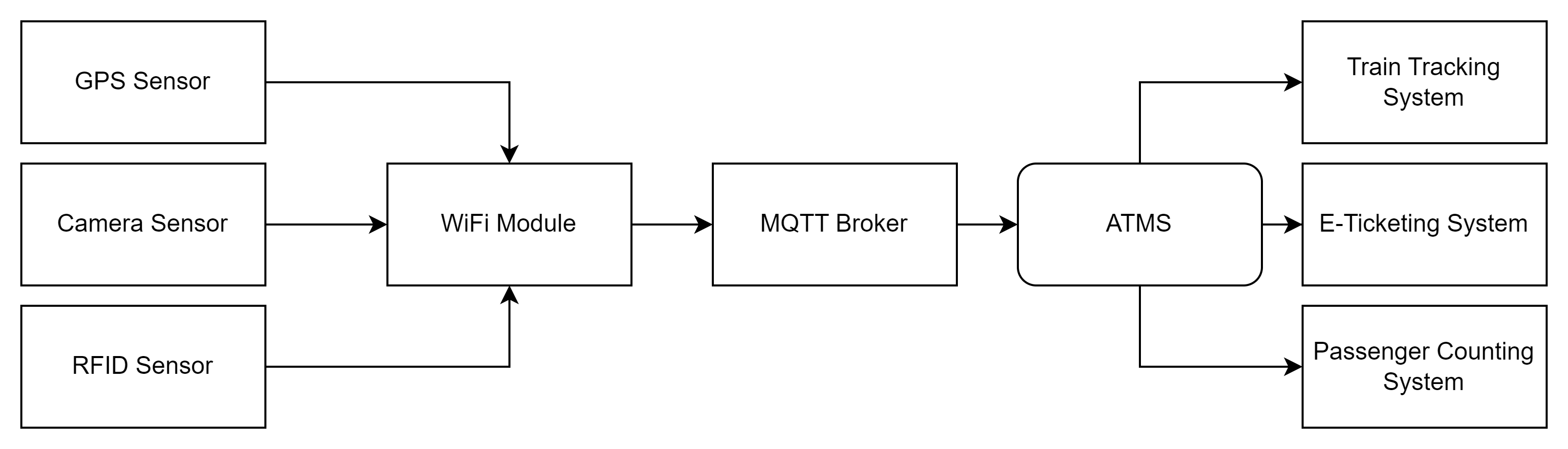}
    \caption{System diagram of the ATMS.}
    \label{fig:atms}
\end{figure}

Figure \ref{fig:atms} illustrates shows the structure of our ATMS system. The entire system is integrated using the MQTT protocol, with infrastructure combined through MQTT nodes and topics.

\subsection{GPS based real time train tracking system with alarm notification system}

\label{sec:s1}
Real-time train tracking with accurate information is crucial for passengers and a key expectation of smart transportation systems. GPS technology is used to track the train's real-time location via a NEO-6M GPS sensor, which captures continuous longitude and latitude data. This sensor is mounted on the train, and the collected location data is published to an MQTT broker. The broker then disseminates this information to a subscribed dashboard for real-time location visualization. For performance comparison, another GPS-based tracking system using the HTTP protocol was developed and evaluated against the MQTT-based GPS tracking system. The evaluation results are presented in Section \ref{sec:hq}. The system employs the Haversine formula \cite{Dauni_2019}, as depicted in Equation \ref{eq:distance}, to calculate the distance to the destination. The distance between the destination and current train location is denoted as $\omega$. The $\theta_1$ and $\theta_2$ are the latitudes of the two points in radians, while $\phi_1$ and $\phi_2$ are the longitudes of the two points in radians. 
 
\begin{equation} \label{eq:distance}
\begin{split}
\omega = & 2 \cdot R \cdot \arcsin \left( \sqrt{\sin^2\left(\frac{{\theta_2 - \theta_1}}{2}\right)} \right. \\
& \left. + \cos(\theta_1) \cdot \cos(\theta_2) \cdot \sin^2\left(\frac{{\phi_2 - \phi_1}}{2}\right) \right)
\end{split}
\end{equation}

The distance to the destination is calculated using the provided equation, and if it exceeds the threshold, the alarm triggers. The equation \ref{eq:distance} calculates the distance between two points. As train tracks are typically straight, the equation gives a reasonably accurate estimate. Passengers can activate this alarm through the app, ensuring control over their journey and avoiding missing their stop. Such a system is vital for stress-free travel, especially on long-distance routes.

\subsection{RFID-based E-ticketing system}
\label{sec:et}

The paper scarcity in Sri Lanka due to high inflation, our RFID-based E-ticketing system offers a timely solution to reduce paper waste. It introduces the concept of a lifelong journey with a single electronic card. Passengers can purchase train tickets using an E-pass, with costs deducted directly from their account, rechargeable for top-ups. Operating on the MQTT protocol, scanned information is published to the MQTT broker and transmitted to our ATMS website, facilitating online seat booking and ticket checking.
\subsection{An vision-based real time people counting system}
\label{sec:pc}

Implementing a people counting system in train transportation enhances safety, efficiency, and service delivery by preventing overcrowding and enabling resource allocation adjustments. The ESP32 camera captures video data, processed by our YOLOv8-based model to identify people entering the train. The counting system calculates detected individuals using a horizontal reference line. The total number of passengers $\lambda_T$ present in a particular compartment, as given in Equation \ref{eq:one}, is calculated based on $\lambda_E$ and $\lambda_O$ while the $\lambda_I$ is denoted as the number of people entering the train and $\lambda_O$ is denoted as those who are exiting from the train. This calculation involves analyzing the passage of individuals across this line to determine the total passenger count within the compartment.

\begin{equation} \label{eq:one}
  \lambda_T = \lambda_I - \lambda_O
\end{equation}

The value of $\lambda_T$ is published using MQTT broker to the ATMS website which acts as the MQTT subscriber. Likewise, the real-time counting information is displayed on the web application.

\subsection{Communication}
\label{sec:com}

Suggesting a standardized MQTT topic for vehicle tracking is crucial for fostering collaboration and interoperability among stakeholders, not only in the railway sector but also in other modes of transportation such as buses.

\begin{figure}[ht]
    \centering
    \includegraphics[width=1\linewidth]{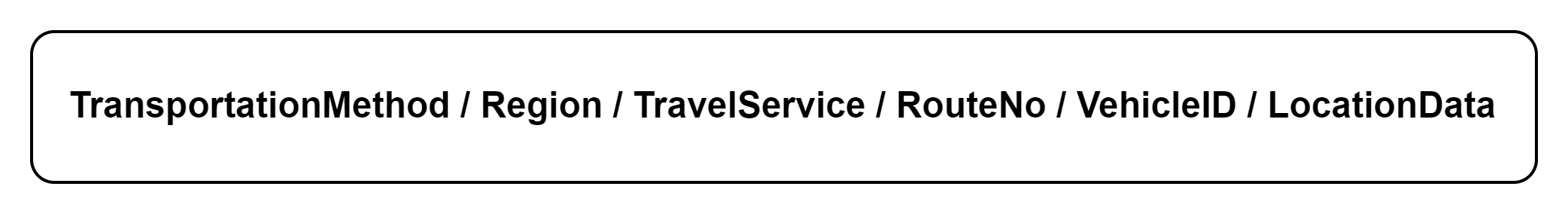}
    \caption{Suggested MQTT topic for the transportation system.}
    \label{fig:topic}
\end{figure}

The hierarchical structure of the topic, encompassing parameters which are stated in the Figure \ref{fig:topic}, ensures a systematic approach to data transmission. The structure allows for flexibility and scalability by incorporating wildcard masks for certain parameters. For instance, the "Method" parameter accommodates various transportation modes such as trains and buses, enabling tracking across different types of vehicles. The "TravelService" parameter utilizes wildcards for categorizing services into long distance or short distance. Additionally, incorporating "Region" as a parameter recognizes the diverse geographic areas and enables connecting the service to the nearest MQTT server. This standardized approach promotes consistency and facilitates seamless integration and data exchange across diverse tracking systems, thereby enhancing operational efficiency and stimulating innovation within transportation management practices.

\subsection{ATMS}
\label{sec:tsm}
The ATMS serves as a centralized system for other subsystems that can be accessible through a user-friendly website for both users and management. The ATMS system is provides a platform for the users to interact with our services which include, train tracking facility, location alarm notification facility, E-ticketing facility. Firstly, users need to register with the system. Once registered, if they wish to access the E-ticketing facility, they can apply for it through our system.

\section{Results and Discussion}
\label{sec:rslt}

\subsection{Evaluation of GPS tracking system}
\label{sec:gps}

An experiment was initiated to verify the accuracy of the neo6m GPS module in producing GPS location information. The longitude and latitude information provided by Google Maps for a specific location was compared with the location data produced by the Neo 6m GPS module for the same location, serving as the basis for evaluating our GPS-based tracking system.

\begin{figure}[ht]
    \centering
    \includegraphics[width=0.7\linewidth]{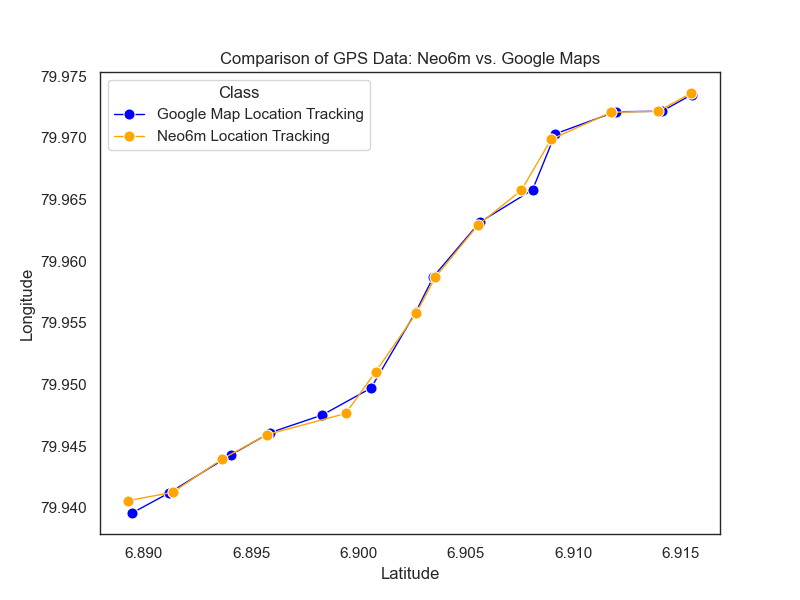}
    \caption{Evaluation of location gathering using Neo 6m GPS module.}
    \label{fig:gpsE}
\end{figure}

As depicted in Figure \ref{fig:gpsE}, the location data produced by the Neo 6m GPS module closely aligns with the actual location. Error distance, denoted by $\Theta$, serves as a metric for verifying and evaluating the system's performance.Equation \ref{eq:distance} is utilized to compute the error between the actual location and the location provided by the GPS module. The average $\Theta$ score for the collected dataset is 24 meters.

\subsection{Evaluation of HTTP and MQTT based GPS tracking system}
\label{sec:hq}

The performance of the MQTT-based GPS tracking system is compared to the HTTP-based system. The MQTT tracker utilizes the Adafruit MQTT broker, while the HTTP tracker operates within Firebase. Both systems are developed and evaluated in Node-RED under identical network conditions to ensure accurate assessment and mitigate potential biases. The average latency for publishing MQTT protocol based location information to the cloud serves as one of the evaluation metric for this experiment, denoted by $\Phi_M$ as mentioned in the equation \ref{eq:em} which is derived from $\Pi_M$ and N. $\Pi_M$ is represents the cumulative latency for the dataset while N is represents the size of the dataset. The average latency for the MQTT based information publishing produced an $\Phi_M$ value of 458.83ms.

\begin{equation} \label{eq:em}
\Phi_M = \frac{\Pi_M}{N}\
\end{equation}

The average latency for HTTP protocol based location information publishing serves as the evaluation metric for HTTP based tracking that is denoted by $\Phi_H$ as mentioned in the equation \ref{eq:eh} which is derived from $\Pi_H$ and N. $\Pi_H$ is represents the cumulative latency of the HTTP based publishing for the dataset.The average latency for the HTTP based information publishing produced an $\Phi_H$ score of 899.5ms. 

\begin{equation} \label{eq:eh}
\Phi_H = \frac{\Pi_H}{N}\
\end{equation}

\begin{figure} [ht]
    \centering
    \includegraphics[width=1\linewidth]{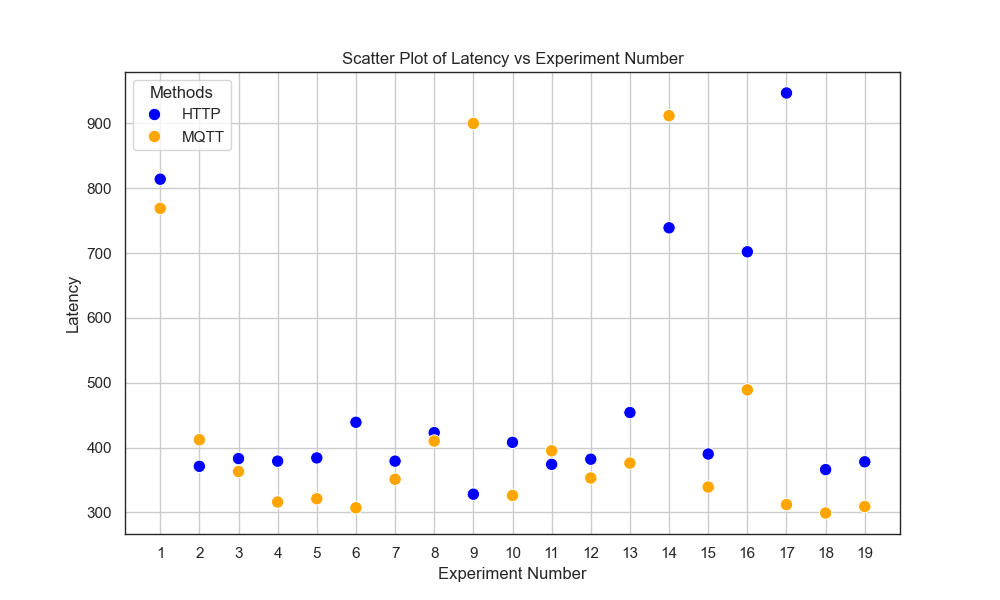}
    \caption{Performance of MQTT over HTTP protocol.}
    \label{fig:hq}
\end{figure}

Figure \ref{fig:hq} shows the performance results of the HTTP and MQTT protocols. The plotted results also depicts the latency measurements of these two protocols over the each experiment.

\subsection{Evaluation of People counting system}
\label{sec:pcs}

\begin{figure}[ht]
    \centering
    \includegraphics[width=0.7\linewidth]{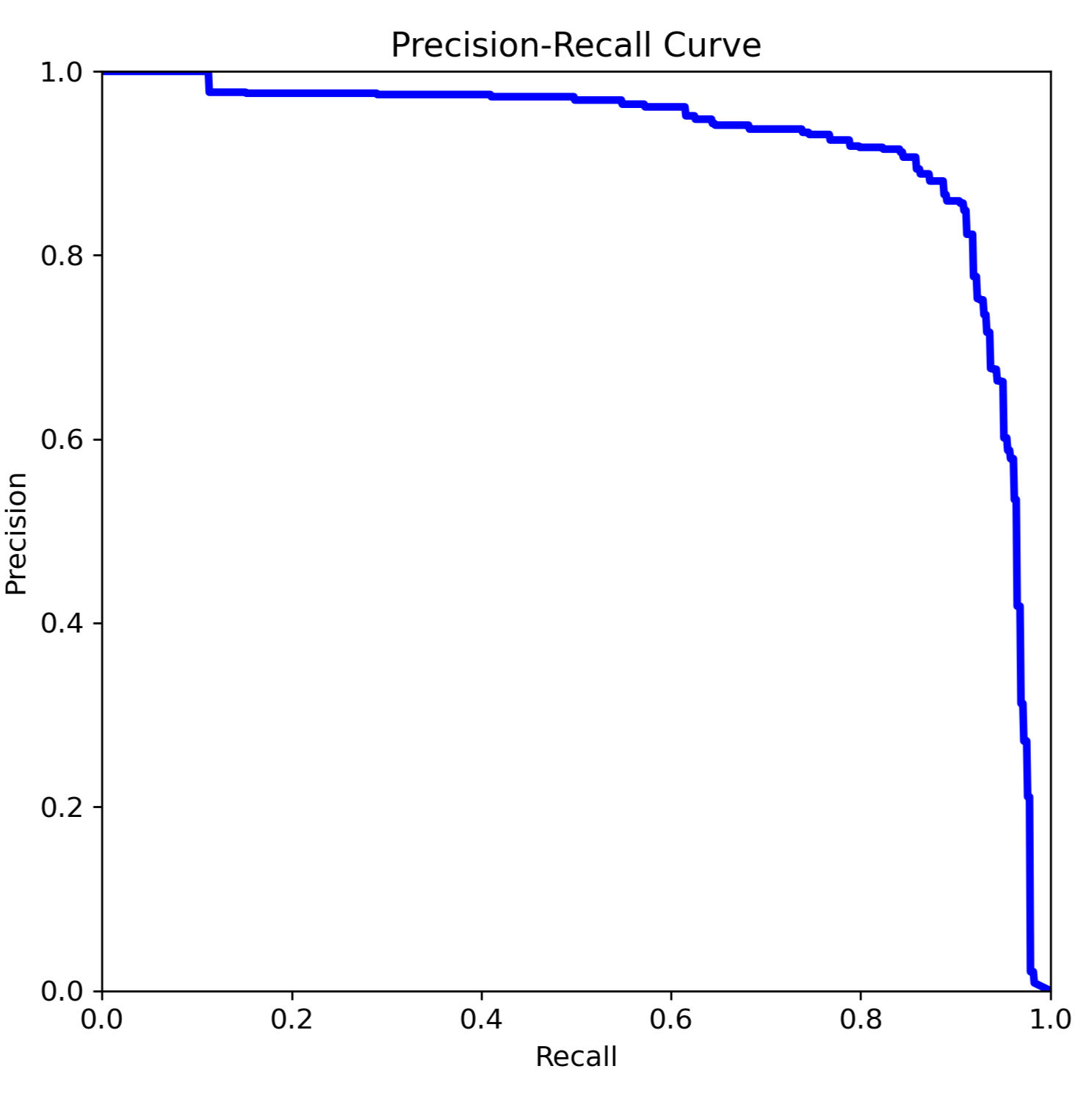}
    \caption{Evaluation of vision based people counting system.}
    \label{fig:psc}
\end{figure}

The precision-recall curve is a crucial metric for evaluating the people detection algorithm's effectiveness, especially with labeled datasets. In Figure \ref{fig:psc}, precision (y-axis) represents accurately predicted positives among all predicted positives, while recall (x-axis) shows the algorithm's ability to identify actual positives in the dataset. The Area Under the Precision-Recall Curve (AUC-PR) validates the algorithm's exceptional performance, confirming its efficacy in real-world applications.

\section{Conclusion}
\label{sec:con}

The proposed system innovatively enhances Sri Lanka's train transportation with a multi-subsystem approach, including a GPS-based real-time tracking system, an RFID-based e-ticketing system, and a vision-based people counting system. A standardized MQTT topic supports system interoperability and integration. The vision-based system uses the YOLOv8 algorithm, showing superior performance with a high AUC-PR value. The GPS-based tracking system, using the Neo 6m module, achieves up to 24 meters accuracy. Utilizing MQTT protocol, the system demonstrated an average latency of 458.83ms, compared to 899.5ms with HTTP, proving MQTT to be twice as fast for GPS-based tracking.

\bibliography{root}

\begin{thebibliography}{10}
\providecommand{\url}[1]{#1}
\csname url@rmstyle\endcsname
\providecommand{\newblock}{\relax}
\providecommand{\bibinfo}[2]{#2}
\providecommand\BIBentrySTDinterwordspacing{\spaceskip=0pt\relax}
\providecommand\BIBentryALTinterwordstretchfactor{4}
\providecommand\BIBentryALTinterwordspacing{\spaceskip=\fontdimen2\font plus
\BIBentryALTinterwordstretchfactor\fontdimen3\font minus \fontdimen4\font\relax}
\providecommand\BIBforeignlanguage[2]{{%
\expandafter\ifx\csname l@#1\endcsname\relax
\typeout{** WARNING: IEEEtran.bst: No hyphenation pattern has been}%
\typeout{** loaded for the language `#1'. Using the pattern for}%
\typeout{** the default language instead.}%
\else
\language=\csname l@#1\endcsname
\fi
#2}}

\bibitem{7814989}
T.~Yokotani and Y.~Sasaki, ``Comparison with http and mqtt on required network resources for iot,'' in \emph{2016 International Conference on Control, Electronics, Renewable Energy and Communications (ICCEREC)}, 2016, pp. 1--6.

\bibitem{8394143}
S.~Saritha and V.~Sarasvathi, ``A study on application layer protocols used in iot,'' in \emph{2017 International Conference on Circuits, Controls, and Communications (CCUBE)}, 2017, pp. 155--159.

\bibitem{7266601}
K.~Radnosrati, F.~Gunnarsson, and F.~Gustafsson, ``New trends in radio network positioning,'' in \emph{2015 18th International Conference on Information Fusion (Fusion)}, 2015, pp. 492--498.

\bibitem{8065840}
M.~Desai and A.~Phadke, ``Internet of things based vehicle monitoring system,'' in \emph{2017 Fourteenth International Conference on Wireless and Optical Communications Networks (WOCN)}, 2017, pp. 1--3.

\bibitem{9683883}
S.~Jawad, H.~Munsif, A.~Azam, A.~H. Ilahi, and S.~Zafar, ``Internet of things-based vehicle tracking and monitoring system,'' in \emph{2021 15th International Conference on Open Source Systems and Technologies (ICOSST)}, 2021, pp. 1--5.

\bibitem{8273112}
M.~B. Yassein, M.~Q. Shatnawi, S.~Aljwarneh, and R.~Al-Hatmi, ``Internet of things: Survey and open issues of mqtt protocol,'' in \emph{2017 International Conference on Engineering \& MIS (ICEMIS)}, 2017, pp. 1--6.

\bibitem{9708053}
A.~C.~M. Nafrees, S.~M.~S. Raseez, C.~G. Ubeshanan, K.~Achutharaj, and A.~L. Hanees, ``Intelligent transportation system using smartphone,'' in \emph{2021 5th International Conference on Electrical, Electronics, Communication, Computer Technologies and Optimization Techniques (ICEECCOT)}, 2021, pp. 229--234.

\bibitem{9946624}
H.~Weligamage, S.~Wijesekara, M.~Chathwara, H.~Isuru~Kavinda, N.~Amarasena, and N.~Gamage, ``An approach of enhancing the quality of public transportation service in sri lanka using iot,'' in \emph{2022 IEEE 13th Annual Information Technology, Electronics and Mobile Communication Conference (IEMCON)}, 2022, pp. 0311--0316.

\bibitem{9501928}
P.~V. Crisgar, P.~R. Wijaya, M.~D.~F. Pakpahan, E.~Y. Syamsuddin, and M.~O. Hasanuddin, ``Gps-based vehicle tracking and theft detection systems using google cloud iot core \& firebase,'' in \emph{2021 International Symposium on Electronics and Smart Devices (ISESD)}, 2021, pp. 1--6.

\bibitem{10277708}
Noerlina, A.~Khairunnisa, and Meiryani, ``Assessment of e-ticketing technology in concert website: A review of benefits, profits, and customer satisfaction,'' in \emph{2023 International Conference on Information Management and Technology (ICIMTech)}, 2023, pp. 1--5.

\bibitem{8537302}
S.~Kazi, M.~Bagasrawala, F.~Shaikh, and A.~Sayyed, ``Smart e-ticketing system for public transport bus,'' in \emph{2018 International Conference on Smart City and Emerging Technology (ICSCET)}, 2018, pp. 1--7.

\bibitem{9732728}
Y.~Punarvit, K.~Sawant, K.~P. k.~R. Shankar, and V.~Kumar, ``Implementation of cashless bus ticketing system using rfid and iot,'' in \emph{2021 International Conference on Advances in Technology, Management \& Education (ICATME)}, 2021, pp. 249--253.

\bibitem{9391951}
E.~P. Myint and M.~M. Sein, ``People detecting and counting system,'' in \emph{2021 IEEE 3rd Global Conference on Life Sciences and Technologies (LifeTech)}, 2021, pp. 289--290.

\bibitem{Chen_2024}
\BIBentryALTinterwordspacing
Y.~Chen, ``The investigation of performance comparison for vgg, yolo, and dino in image classification,'' \emph{Highlights in Science, Engineering and Technology}, vol.~85, p. 984–990, Mar. 2024. [Online]. Available: \url{https://drpress.org/ojs/index.php/HSET/article/view/18546}
\BIBentrySTDinterwordspacing

\bibitem{Dauni_2019}
\BIBentryALTinterwordspacing
P.~Dauni, M.~D. Firdaus, R.~Asfariani, M.~I.~N. Saputra, A.~A. Hidayat, and W.~B. Zulfikar, ``Implementation of haversine formula for school location tracking,'' \emph{Journal of Physics: Conference Series}, vol. 1402, no.~7, p. 077028, dec 2019. [Online]. Available: \url{https://dx.doi.org/10.1088/1742-6596/1402/7/077028}
\BIBentrySTDinterwordspacing

\end{thebibliography}
\end{document}